\documentclass[journal,12pt,onecolumn]{IEEEtran}
\usepackage{color}
\usepackage{verbatim}

\usepackage{ifpdf}

\usepackage{cite}

\usepackage{subcaption}

\ifCLASSINFOpdf
   \usepackage[pdftex]{graphicx}
   \graphicspath{{./TAC14/pdf}{./Figures/}}
   \DeclareGraphicsExtensions{.pdf,.jpeg,.png}
\else
   \usepackage[dvips]{graphicx}
   \graphicspath \graphicspath{{./TAC/eps}{./Figures/}}
   \DeclareGraphicsExtensions{.eps}
\fi
\usepackage{overpic}

\usepackage[cmex10]{amsmath}

\usepackage{amssymb}
\usepackage{mathtools}

\usepackage{url}

\newtheorem{Thm}{Theorem}

\newtheorem{Lemma}{Lemma}

\begin{document}
%
\title{Bounded Disturbance Amplification for Mass Chains with Passive Interconnection}
%
%
%

\author{Kaoru~Yamamoto~and~Malcolm~C.~Smith
\thanks{This work was supported by the Funai Overseas Scholarship.}
\thanks{K.~Yamamoto and M.~C.~Smith are with the Department of Engineering, University of Cambridge,
Cambridge CB2 1PZ, U.K. (email: ky255@cam.ac.uk and mcs@eng.cam.ac.uk).}}
\maketitle

\begin{abstract}
This paper introduces the problem of passive control of a chain of $N$
identical masses in which there is an identical passive connection between
neighbouring masses and a similar connection to a movable point. The
problem arises in the design of multi-storey buildings which are subjected to
earthquake disturbances, but applies in other situations, for example vehicle
platoons. The paper studies the scalar transfer functions from the disturbance
to a given intermass displacement. It is shown that these transfer functions
can be conveniently represented in the form of complex iterative maps and that
these maps provide a method to establish boundedness in $N$ of the $\mathcal{H}_\infty$-norm
of these transfer functions for certain choices of interconnection impedance.
\end{abstract}



\section{Introduction}
\IEEEPARstart{T}{he} main focus in this paper is disturbance amplification in a
chain of masses with passive interconnection. The problem is motivated by the
problem of vibration suppression in multi-storey buildings subjected to earthquake
disturbances. One of the main objectives for seismic design is to limit the inter-storey
displacements in response to disturbances. For this purpose, the installation of passive
control devices between floors is widely accepted \cite{Soo97, Con98, Tak09}. In the present paper we
consider passive interconnections of the most general type, which may require the use
of inerters \cite{Smi02} in addition to springs and dampers. The use of such devices is
already under consideration for multi-storey buildings \cite{Wan07, Wan10, Laz14}.

With an increasing trend to build ever taller buildings, the general question arises whether
it is possible to achieve uniform boundedness of disturbance amplification as the number
of storeys grows. This can be characterised as a ``scalability'' property. More precisely in the
present context, this is the question whether the $\mathcal{H}_\infty$-norm of the frequency responses from
ground disturbance to the individual inter-storey displacements remains bounded as the
number of storeys increases.

A similar problem has been considered in the literature on automatic control of vehicles; see
for example \cite{Swa96, Sei04, Bar05, Les07, Mid10}. Using terminology from that area, our
problem formulation corresponds to ``symmetric bidirectional control'', i.e., control laws in which
the control action for each vehicle is equally dependent on the spacing errors with the predecessor
and the follower. In \cite{Bar05} a general result has been shown that, using symmetric bidirectional control,
the infinity norm of the transfer function vector from lead vehicle trajectory to spacing error grows
without bound as the number of vehicles increases, if the combined vehicle-controller dynamics
contains a double integrator. This corresponds to a positive static (spring) stiffness in the case of a mass
chain with passive interconnection, which is the usual case. However, it may be noted
that well-regulated individual intermass displacements may be a satisfactory performance
objective for buildings.
In \cite{Jov05} it is pointed out that, for a large vehicular platoon, the least stable
closed-loop pole tends to the origin as the size of platoon increases, and hence, the time constant of the closed-loop
system grows without bound. It is also interesting that, in \cite{Bam12}, it is observed that slow accordion-like motion
of the entire formation  in a large vehicular platoon may not be inconsistent with the spacing between each vehicle
being well regulated.

\begin{figure}[!bt]
      \centering
      \includegraphics[trim=70mm 120mm 120mm 50mm, clip, scale=0.5]{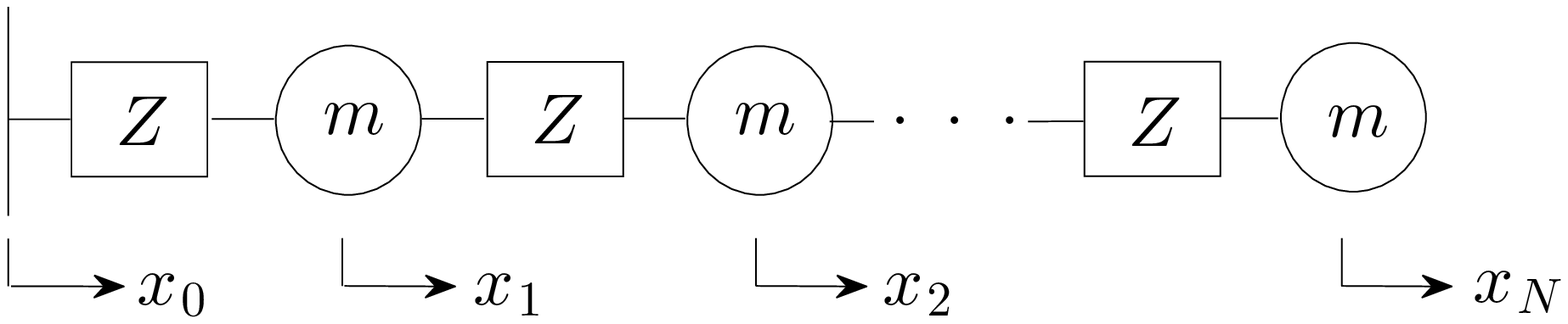}
      \caption{Chain of $N$ masses $m$ connected by a passive mechanical impedance $Z(s)$
      (admittance $Y(s)=Z(s)^{-1}$), and connected to a movable point $x_0$.}
      \label{fig:Mchains}
   \end{figure}

In the present paper we study the scalar transfer functions from the movable point displacement $x_0$ to a given individual
intermass displacement  (an inter-storey drift in the building application) in a chain of $N$ identical masses with identical
passive interconnection (Fig.~\ref{fig:Mchains}). It is shown that these transfer functions can be defined recursively in $N$.
These recursions give a convenient method to accurately compute these transfer functions. They can also be
interpreted as complex iterative maps, in particular, iterated M\"{o}bius transformations. Making use of the properties
of M\"{o}bius transformations, the fixed points of these recursions are shown to provide the asymptotic behaviours
of these transfer functions. It is also shown that the $\mathcal{H}_\infty$-norm of these individual transfer functions is bounded
above independently of the length of the mass chain for a suitable choice of the interconnection impedance.
The paper goes further to provide a graphical means to design a suitable interconnection impedance so that
the supremum of the $\mathcal{H}_\infty$-norm over $N$ is no greater than a prescribed value. This can be thought
of as an $\mathcal{H}_\infty$ control design for an infinite family of plants in which the interconnection impedance
is the controller.

The paper is structured as follows. In Section~\ref{sec:passivity}, we present some definitions and facts on passive
mechanical networks.  In Section~\ref{sec:probf}, a mass chain model with passive interconnections is introduced.
We describe the transfer functions from a movable point to a given intermass displacements as a function of a
dimensionless parameter $h$ depending on the impedance and mass. Then the stability of the system is discussed
in Section~\ref{sec:stability}. Section~\ref{sec:main} is the main part of the paper. The transfer functions are described
in the form of iterated M\"{o}bius transformations and the asymptotic behaviours are discussed.
Theorem~\ref{thm:scalability} shows the boundedness result. These results are also illustrated graphically
in this section. In numerical examples, we compare a standard spring-damper suspension to the use of inerters.
Some concluding remarks follow in Section~\ref{sec:concl}.
\section{Preliminaries}
\subsection{Background on Passive Mechanical Networks} \label{sec:passivity}
A mechanical one-port network with force-velocity pair $(F, v)$ is {\em passive\/} if for all
square integrable pairs $F(t)$ and $v(t)$ on $(-\infty,T]$, $\int_{-\infty}^T F(t)v(t) dt \geq 0$ \cite{And06}.
For a linear time-invariant network the impedance $Z(s)$ is defined by the ratio
$\hat{v}(s)/\hat{F}(s)$ where $\hat{}$ denotes the Laplace transform, and $Y(s)=Z(s)^{-1}$ is
called the admittance. Such a network can be shown to be passive if and only if $Z(s)$ or $Y(s)$
is positive real \cite{Bru31,Val60}. A real-rational function $G(s)$ is {\em positive real} if $G(s)$
is analytic and $\operatorname{Re}(G(s)) \geq 0$ in $\operatorname{Re}(s)>0$.

The passive components considered are springs, dampers and inerters. The {\em inerter} is
a mechanical two-terminal, one-port device with the property that the applied force at the terminals
is proportional to the relative acceleration between the terminals, i.e., $F=b(\dot{v}_2-\dot{v}_1)$
where $b$ is the constant of proportionality called the {\em inertance} which has units of kilograms
\cite{Smi02} and $v_1, v_2$ are the terminal velocities. The inerter completes a standard analogy
between mechanical and electrical networks which allows classical results from electrical network
synthesis to be translated over exactly to mechanical systems. In particular, any real-rational
positive-real function can be realised as the impedance or admittance of a network with springs,
inerters, and dampers only \cite{Smi02}.
\section{Problem Formulation}\label{sec:probf}
\subsection{General notation}
The set of natural, real and complex numbers is denoted by $\mathbb{N}$, $\mathbb{R}$, $\mathbb{C}$, respectively.
$\mathbb{R}^{m \times n}$ is the set of $m$ by $n$ real matrices. $\mathbb{R}_+$ is the set of non-negative
numbers and $\mathbb{C}_+$ is the closed right-half plane. $\mathcal{H}_\infty$ is the standard Hardy space on the
right-half plane and $\left\|{\cdotp}\right\|_\infty$ represents the $\mathcal{H}_\infty$-norm. The composition of two
functions is denoted by $f \circ g (x) = f(g(x)) $ and $f^n(x)$ is the $n$-fold composition of $f$.
\subsection{Chain model}
We consider a chain of $N$ identical masses $m$ connected by identical passive mechanical
networks (Fig.~\ref{fig:Mchains}). Each passive mechanical network provides an equal and opposite
force on each mass and is assumed here to have negligible mass. The system is excited by a movable
point $x_0(t)$ and the displacement of the $i$th mass is denoted by $x_i(t)$, $i \in \{1,2, \dots, N\}$.
We assume that the initial conditions of the movable point and the mass displacements are all zero.

The equations of motion in the Laplace transformed domain are
\begin{align*}
  ms^2 \hat{x}_i &= sY(s)( \hat{x}_{i-1} - \hat{x}_i)+ sY(s)(\hat{x}_{i+1}- \hat{x}_i) \quad\mbox{for } i=1,\dots,N-1, \\
  ms^2 \hat{x}_N &= sY(s)( \hat{x}_{N-1} - \hat{x}_N),
\end{align*}
where $\hat{}$ denotes the Laplace transform.
In matrix form this can be written as
\begin{equation*}
 ms^2 \hat{x} = sY(s)H_N \hat{x} + sY(s)e_1 \hat{x}_0
\end{equation*}
and hence
\begin{equation}
\hat{x} = (h(s)I_N-H_N)^{-1}e_1\hat{x}_0 \label{eq:solvex}
\end{equation}
where
$I_N$ is the $N \times N$ identity matrix,
\begin{align*}
 & h(s) = sZ(s)m, \quad Z = Y^{-1}, \\[0.5em]
 & \hat{x} = [\hat{x}_1, \dots,  \hat{x}_N]^{\mathrm{T}},\quad e_1 = [1,\ 0,\dots,\ 0]^{\mathrm{T}} \in \mathbb{R}^N, \\[0.5em]
 & H_N= \begin{bmatrix}
         -2    & 1      & 0      & \cdots & 0      \\
         1     & -2     & 1      & \ddots & \vdots \\
         0     & \ddots & \ddots & \ddots & 0      \\
         \vdots& \ddots & 1      & -2     & 1      \\
         0     & \cdots & 0      & 1      & -1     \\
        \end{bmatrix}\in \mathbb{R}^{N\times N}. \nonumber
 \end{align*}

Let us consider the characteristic polynomials $d_i$ of ${H_i\in\mathbb{R}^{i\times i}}$ in the variable $h$
 given by
\begin{eqnarray}
 d_i &=& \mathrm{det}{(hI_i-H_i)} \nonumber \\
     &=& \begin{vmatrix}
         h+2    & -1      & 0       & \dots  & 0      \\
         -1     & h+2     & -1      & \ddots  & \vdots \\
         0      & \ddots  & \ddots  & \ddots  & 0      \\
         \vdots & \ddots  & -1      & h+2     & -1      \\
         0      & \dots  & 0       & -1      & h+1     \\
        \end{vmatrix}
\quad\mbox{for } i=1,\dots,N \label{eq:dn}.
\end{eqnarray}
Then $d_1=h+1$. Suppose also $d_{-1} = 1$ and $d_0 = 1$. Using the Laplace expansion of \eqref{eq:dn}, we find that
\begin{equation}
  d_i(h) = (h+2)d_{i-1}(h) -d_{i-2}(h) \quad \mbox{for } i=1,\dots,N. \label{eq:dnrec}
\end{equation}
Equation \eqref{eq:solvex} can be written using $d_i$ as
\begin{eqnarray}
 \hat{x} &=& \frac{\mathrm{adj}{(h(s)I_N-H_N)}}{\mathrm{det}{(h(s)I_N-H_N)}}e_1 \hat{x}_0  \nonumber \\
       &=& \frac{1}{d_N}
                  \begin{bmatrix}
                    d_{N-1} & *      & \cdots \\
                    \vdots  & \vdots &        \\
                    d_0     & *      & \cdots \\
                 \end{bmatrix}
                  \begin{bmatrix}
                    1\\
                    0\\
                    \vdots\\
                    0\\
                    \end{bmatrix}\hat{x}_0    \nonumber \\
          &=&     \begin{bmatrix}
                    d_{N-1}/d_N\\
                    \vdots\\
                    d_0/d_N\\
                    \end{bmatrix} \hat{x}_0 \label{eq:disp}.
 \end{eqnarray}
Then the intermass displacement of the $i$th mass defined by $\delta_i = x_i - x_{i-1}$
in the Laplace domain is given by
\begin{equation}
 \hat{\delta_i} =\left(\left(d_{N-i}-d_{N-i+1}\right)/{d_N}\right)\hat{x}_0 =: T_{\!{\hat{x}_0\rightarrow\hat{\delta}_i}} \hat{x}_0  \label{eq:TF}
\end{equation}
for $i=1, \dots, N$.
\section{Stability of Passive Interconnection} \label{sec:stability}
We first establish some properties of the sequence $d_{i}(h)$, treating $h$ as the independent variable,
namely that they are Hurwitz with real distinct roots in the interval $(-4, 0)$ for $i=1, 2, \dots$ and
form a Sturm sequence.
\begin{Thm} \label{thm:di}
~
\begin{enumerate}
 \item $d_i(h)$ has negative real distinct roots which interlace the roots of $d_{i+1}(h)$ for
  $i=1, 2, \dots$.
  \item The roots of $d_i(h)$ lie in the interval $(-4, 0)$ for $i=1, 2, \dots$.
\end{enumerate}
\end{Thm}
\begin{IEEEproof}
 \begin{enumerate}
   \item  It is evident that $d_i(0)=1$ and $d_i(h)$ are continuous and monic for all $i$.
          Let ${a_m} \mbox{ for } m=1,\dots, n+1$ denote the roots of $d_{n+1}(h)$ and
         ${b_m}$ the roots of $d_n(h) \mbox{ for } m=1,\dots, n$.
         Suppose the result holds for the $i=n$, namely, $0>a_1>b_1>a_2>b_2>\cdots>a_n>b_n>a_{n+1}$.
         Since $d_{n+1}(0)=1$ and $d_{n+2}(a_1)=-d_n(a_1)<0$, $d_{n+2}(h)$ has at least one root
         in $(a_1, 0)$. Similarly, $d_{n+2}(h)$ has at least one root in each interval $(a_m, a_{m-1})
         \mbox{ for } m=2,\dots, n+1$ since $d_{n+2}(a_{m-1})d_{n+2}(a_m)=(-d_n(a_{m-1}))(-d_n(a_m))<0$
         using \eqref{eq:dnrec}.
         Further $d_n(h)$ and $d_{n+2}(h)$ have the same sign in the limit as $h \rightarrow -\infty$, which is
         opposite to that of $d_{n+1}(h)$. This implies there exists at least one root of $d_{n+2}(h)$
         in $(-\infty, a_{n+1} )$. Since $d_{n+2}(h)$ has at most $n+2$ roots, it has exactly one root in each interval.
         Hence the result holds for $i=n+1$. It is straightforward to check the case of $i=1$, and the proof
         then follows by induction.

   \item Let $\mathbb{P}=\cup _N {\sigma(H_N)}$ where ${\sigma(\cdotp)}$ denotes the spectrum. Note that
         a Gershgorin disc bound on the eigenvalues of $H_N$ \cite{Bar05,Hor99,Les07} (this holds for all $N$) implies
         ${\mathbb{P}}\subset [-4,0]$. It is straightforward to check that $d_i(0)=1$ and $d_i(-4)=(-1)^i(2i+1)$. Hence,
         the roots of $d_i(h)$ lie in the interval $(-4,0)$. 
 \end{enumerate}
 \end{IEEEproof}
We will say that the system of Fig.~\ref{fig:Mchains} is stable if all poles in the transfer functions $T_{\!{\hat{x}_0\rightarrow\hat{\delta}_i}}$ have negative
real parts (in the $s$-domain). We note that \cite{Har14} has investigated the stability of systems which are a generalised
version of our model using the notion of ``generalized frequency variables.'' Here we provide an explicit condition for stability
for a general $N$.
\begin{Thm}\label{thm:stability}
For $0\not\equiv Z(s)$ positive real, the system of Fig.~\ref{fig:Mchains} is stable if $sZ(s)m$ does not take values in the
interval $(-4, 0)$ for any $s$ with $\operatorname{Re}(s)=0$.
\end{Thm}
\begin{IEEEproof}
From \eqref{eq:disp}, poles in $T_{\!{\hat{x}_0\rightarrow\hat{\delta}_i}}$ can only occur at an $s$ for which $d_N(h(s))=0$. From \cite{Bru31} (Theorem~VI)
$\operatorname{Re}(Z(s)) > 0$ for $\operatorname{Re}(s) > 0$. The result now follows from Theorem~\ref{thm:di}.
\end{IEEEproof}
\section{Intermass Displacements} \label{sec:main}
It is shown in this section that the transfer functions from the disturbance to a given intermass displacement in a chain
of $N$ masses are represented in the form of complex iterative maps.
\begin{Thm} \label{thm:ithrec}
For any $i=1, 2, \dots$, intermass displacements in a chain of $N$ masses satisfy the recursion:
\begin{equation}
-T_{\!{\hat{x}_0\rightarrow\hat{\delta}_i}}=:F_N^{(i)} = \frac{d_{i-2}F^{(i)}_{N-1}+h}{F^{(i)}_{N-1}+d_i} \label{eq:ithrec}
\end{equation}
for $N= i, i+1, \dots$, where $T_{\!{\hat{x}_0\rightarrow\hat{\delta}_i}}$ is the transfer function from the disturbance
$x_0$ to the $i$th intermass displacement
$\delta_i$, $F^{(i)}_{i-1}=0$, $h(s)=sZ(s)m$ and $d_i$ is as defined in \eqref{eq:dn}.
\end{Thm}
\begin{IEEEproof}
See Appendix~\ref{ap:thm:ithrec}.
\end{IEEEproof}
The above recursion describes a sequence of transfer functions in the complex variable $s$. It can also be interpreted
as a complex iterative map \cite{Dev89} for a given fixed $s \in \mathbb{C}$, or equivalently a fixed $h \in \mathbb{C}$.
In particular, writing
\begin{equation}\label{eq:ithmob}
f_i(z)=\frac{d_{i-2}z+h}{z+d_i}
\end{equation}
we see that the sequence $F_N^{(i)}$ for $N=i-1, i, i+1, \dots$ is the same as $0, f_i(0), f_i(f_i(0)), \dots$ for a given
$h \in \mathbb{C}$.
This is called the orbit of $0$ for the recursion (complex iterative map) defined by \eqref{eq:ithmob}.
\subsection{Convergence to Fixed Points}
A complex number $\mu$ is called a fixed point of a mapping $f$ if $f(\mu)=\mu$. For a fixed $h \in \mathbb{C}$,
the sequence $\{F_N^{(i)}\}$ in \eqref{eq:ithrec} has at most two fixed points $\mu=\mu_\pm^{(i)}$ which satisfy
\begin{equation} \label{eq:quadithmu}
{\mu}^2+(d_i - d_{i-2})\mu-h=0.
\end{equation}
It may be observed that \eqref{eq:ithmob} takes the form of a M\"{o}bius transformation
which has the normalised form \cite{Nee97}
\begin{equation*}
f_i(z) = \frac{az+b}{cz+d}
\end{equation*}
where $a=d_{i-2}/d_{i-1}$, $b=h/d_{i-1}$, $c=1/d_{i-1}$, $d=d_i/d_{i-1}$ and $ad-bc=1$
since
\begin{equation} \label{eq:disquared}
d_{i-2}d_{i}-d_{i-1}^2=h
\end{equation}
which is easily shown by induction.
The properties of the recursion are then determined by $\mbox{trace}^2(f_i)=(a+d)^2=(h+2)^2$ as follows:
$f_i$ is (i) parabolic when $h=0$ or $-4$, (ii) elliptic when $h\in(-4, 0)$ and (iii) loxodromic when $h \not\in [-4, 0]$.
(This uses the terminology of \cite{Bea01} in which hyperbolic maps are a subclass of loxodromic maps.)
The following theorem can be shown by the use of a conjugacy transformation (see \cite{Bea01}).
\begin{Thm}\label{thm:convergence}
\begin{enumerate}
\item When $h=0$ or $-4$, there is a unique fixed point, in this case $\mu_+^{(i)}$, and the sequence $\{F_N^{(i)}\}$
defined by \eqref{eq:ithrec} converges pointwise for any initial condition.
\item When $h \in (-4, 0)$, $\{F_N^{(i)}\}$ fails to converge for any initial condition other than the fixed points.
\item When $h\not\in[-4, 0]$, there are two fixed points, an attractive fixed point and a repulsive fixed point, in
this case $\mu_+^{(i)}$ and $\mu_-^{(i)}$ respectively, and $\{F_N^{(i)}\}$ converges pointwise to $\mu_+^{(i)}$
for any initial condition other than $\mu_-^{(i)}$.
\end{enumerate}\hfill{\IEEEQEDopen}
\end{Thm}

For the specific case of the orbit of $0$, Theorem~\ref{thm:convergence} specialises to: $\{F_N^{(i)}\}$
converges to $\mu_+^{(i)}$ when $h\not\in(-4, 0)$ but fails to converge otherwise. Hence, if
$h(s)\not\in(-4, 0)$ for all $s\in\mathbb{C}_+$,
\[
\sup_{\omega}\lim_{N\rightarrow \infty}|{F_N^{(i)}(h(j\omega))}|=\sup_{\omega}|\mu_+^{(i)}(h(j\omega))|.
\]
Furthermore, $\sup_{\omega} |\mu_+^{(i)}(h(j\omega))| \leq \sup_N \|{F_N^{(i)}(h(s))}\|_\infty$
\big(since $|\mu_+^{(i)}(h(j\omega))| =\lim_{N\rightarrow \infty}|{F_N^{(i)}(h(j\omega))}|$\big).
In Theorem~\ref{thm:absmu} it is shown that $|\mu_+^{(i)}(h)| < 2$ for any $h \not \in [-4, 0]$.
However, it is not clear whether $\sup_N\|F_N^{(i)}(h(s))\|_\infty$ can be suitably bounded or indeed whether it is finite.
This is our main result which will be shown in Theorem~\ref{thm:scalability}. The proof relies on the conjugacy
transformation of $f_i(z)$ which is explicitly described in the next theorem.
\begin{Thm} \label{thm:conjmob}
For $h \not\in [-4, 0]$
\begin{equation} \label{eq:conjmob}
f_i(z)=\frac{d_{i-2}z+h}{z+d_i}=\varphi_i^{-1}\circ\lambda_i\circ\varphi_i(z)
\end{equation}
with
\setcounter{equation}{9}
\begin{subequations}
\begin{align}
\varphi_i(z)&=\frac{z - \mu_+^{(i)}}{z-\mu_-^{(i)}}, \nonumber\\
\varphi_i^{-1}(z)&=\frac{\mu_+^{(i)} - z\mu^{(i)}_{-}}{1-z}, \nonumber\\
\lambda_i(z)&=\frac{d_{i-2}-\mu_+^{(i)}}{d_{i-2}-\mu_-^{(i)}}z \label{eq:conjtrans}\\
		    &=\zeta^2 z \label{eq:multiplier}
\end{align}
where
\begin{equation}\label{eq:zeta}
\zeta=\frac{d_{i-2}-\mu_+^{(i)}}{d_{i-1}}.
\end{equation}
Moreover, $\zeta$ is independent of $i$ and is the root of
\begin{equation} \label{eq:zetaquad}
\zeta^2-(h+2){\zeta}+1 = 0
\end{equation}
satisfying $|\zeta|<1$.
\end{subequations}
\end{Thm}
\begin{IEEEproof}
\eqref{eq:conjmob} with \eqref{eq:conjtrans} follows by direct algebraic computation. In Appendix~\ref{ap:zetaellipse} it is shown that
\eqref{eq:multiplier} and \eqref{eq:zetaquad} hold for $\zeta$ as defined in \eqref{eq:zeta}. We can check directly that
\eqref{eq:zetaquad} has roots $\zeta_+$, $\zeta_-$ satisfying $|\zeta_+|<1<|\zeta_-|$ if and only if $h \not\in [-4, 0]$.
In this case \eqref{eq:zeta} holds with $\zeta=\zeta_+$.
\end{IEEEproof}
We remark that, for $h\not\in[-4, 0]$, the labelling of the roots $\mu_+^{(i)}, \mu_-^{(i)}$ of \eqref{eq:quadithmu}
can be determined by finding the root $\zeta_+$ of \eqref{eq:zetaquad} which satisfies $\zeta_+ < 1$ and then solving \eqref{eq:zeta}
for $\mu_+^{(i)}$.
\begin{Thm} \label{thm:absmu}
For $h\not\in[-4, 0]$
\begin{equation} \label{eq:murec}
\mu_+^{(i+1)}=\zeta\mu_+^{(i)}
\end{equation}
and
\begin{equation}\label{eq:absmu}
|\mu_+^{(i)}| < |\mu_+^{(i-1)}| < \dots < |\mu_+^{(1)}| < 2
\end{equation}
where $\zeta$ is as defined in \eqref{eq:zeta}.
\end{Thm}
\begin{IEEEproof}
Since $\zeta$ is independent of $i$,
\begin{equation*}
\zeta = \frac{d_{i-2} - \mu_+^{(i)}}{d_{i-1}} = \frac{d_{i-1} - \mu_+^{(i+1)}}{d_i}.
\end{equation*}
Therefore,
\begin{align*}
\mu_+^{(i+1)} &=\frac{d_i\mu_+^{(i)}+d_{i-1}^2 - d_{i-2}d_i}{d_{i-1}} \\
&= \frac{d_i\mu_+^{(i)}-h}{d_{i-1}} \qquad\mbox{\big(see \eqref{eq:disquared}\big)}  \\
&= \frac{d_i\mu_+^{(i)}-({\mu_+^{(i)}}^2+(d_i-d_{i-2})\mu_+^{(i)})}{d_{i-1}} \qquad\mbox{\big(see \eqref{eq:quadithmu}\big)}  \\
&=  \frac{d_{i-2} - \mu_+^{(i)}}{d_{i-1}}\mu_+^{(i)} = \zeta\mu_+^{(i)}.
\end{align*}
Since $|\zeta| < 1$ if $h\not\in[-4, 0]$ and $\mu_+^{(1)}=1-\zeta$ given by substituting $i=1$ in \eqref{eq:zeta},
\begin{equation*}
|\mu_+^{(i)}| < |\mu_+^{(i-1)}| < \dots < |\mu_+^{(1)}| =|1-\zeta| < 2.
\end{equation*}
\end{IEEEproof}
\hspace{0.8em} {\em Remark:}
We remark that \eqref{eq:murec} holds also for $h\in[-4, 0]$. In particular, if $h=0$ or $-4$,
\eqref{eq:zetaquad} has a multiple root and \eqref{eq:zeta} determines a unique fixed point.  If $h\in(-4, 0)$, two roots of
\eqref{eq:zetaquad} $\zeta_+$ and $\zeta_-$ satisfy $|\zeta_+|=|\zeta_-|=1$, and either root may be selected for $\zeta_+$
with $\mu_+^{(i)}$ then determined by \eqref{eq:zeta}. Consequently, if $h\in[-4, 0]$,
$|\mu_+^{(i)}| = |\mu_+^{(i-1)}| = \dots = |\mu_+^{(1)}| \leq 2$.
The equality $ |\mu_+^{(1)}| = 2$ holds only when $\zeta=-1$ corresponding to $h=-4$.

\subsection{Bounds on Iterative Maps}

Considering the orbit of $0$ for \eqref{eq:conjmob}, for $N=i-1, i, \dots$,
\begin{align}
F_N^{(i)}&=f_i^{N-i+1}(0)=\varphi_i^{-1}\circ\lambda_i^{N-i+1}\circ\varphi_i(0) \nonumber \\
 &=\mu_+^{(i)}\displaystyle\frac{1-\zeta^{2(N-i+1)}}{1-\displaystyle\frac{\mu_+^{(i)}}{\mu_-^{(i)}}\zeta^{2(N-i+1)}} \nonumber\\
 &=\mu_+^{(i)}\displaystyle\frac{1-\zeta^{2{(N-i+1)}}}{1+\zeta^{2N+1}}\label{eq:mobF}
 \end{align}
since $\varphi_i(0)=\mu_+^{(i)} / \mu_-^{(i)} = -\zeta^{2i-1}$ (see Appendix~\ref{ap:mupovermun}).

We now make use of \eqref{eq:mobF} to establish upper bounds on $| F_N ^{(i)}(h(s))|$ for suitable choices of $h(s)$.
\begin{Thm} \label{thm:scalability}
Suppose $Z(s)=\left(k/s+Y_1(s)\right)^{-1}$ where $k$ is a positive constant and $Y_1(s)$ is a positive-real
admittance satisfying $Y_1(0)>0$. Suppose $h(j\omega)=mj\omega Z(j\omega)$ does not intersect the interval
$[-4, 0)$ for any $\omega \geq 0$.
Then
\begin{equation*}
\sup_{N\geq i} {\left\|F_N^{(i)}\left(h(s)\right)\right\|_\infty}
\end{equation*}
is finite for any $i=1, 2, \dots$.
\end{Thm}
\begin{IEEEproof}
See Appendix~\ref{ap:scalability}.
\end{IEEEproof}
Since $Z(s)$ is positive real \big($\operatorname{Re}(Z(j\omega))\geq 0$ for all $\omega$\big), we note that the condition that $h(s)$ does not
intersect ${[{-4}, 0)}$ is equivalent to $m\omega Z(j\omega)$ not touching the imaginary axis between $(0, j4]$.
This essentially means that the mechanical impedance does not behave in a purely lossless manner
for any frequencies for which $m\omega Z(j\omega) \in (0, j4]$, which is a very mild condition that is easy to satisfy
(and hard to violate) in practice. We note that the condition $Y_1(0) > 0$ can be interpreted in the same manner.

Theorem~\ref{thm:scalability} shows that the individual transfer functions from $x_0$ to
a given intermass displacement are uniformly bounded with respect to the size of the chain of masses for a suitable
choice of $h$.
It is evident that the increasing length of the error vector as $N \rightarrow \infty$ is playing a role
in the unboundedness property of \cite{Bar05}, and that the unboundedness of the vector need not imply that the $\mathcal{H}_\infty$-norm of individual
entries is unbounded with $N$. In this sense our result could be viewed as a relaxation of the definition of string stability.
We point out that \cite{Kno14} has considered a different alternative to string stability, formulated in the time domain, and shown that this may be satisfied by a vehicle string with integral action.
\subsection{Examples}
For the purpose of graphical representations we now introduce the inverse of $h$:
\begin{equation}
g(s)=h^{-1}(s)=Y(s)/(sm). \label{eq:defg}
\end{equation}

\begin{figure}[!tb]
  \centering
     \includegraphics[trim=0mm 10mm 0mm 10mm, clip, scale=0.3]{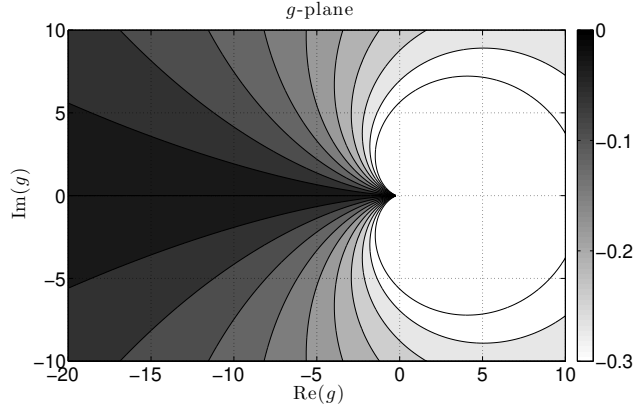}
    \caption{Contour plot of $|\zeta(h)|$ where $h=g^{-1}$.}
    \label{fig:zeta}
\end{figure}

From \eqref{eq:mobF} the speed of convergence of $F_N^{(i)}$ to $\mu_+^{(i)}$ is determined by $|\zeta|$, with the
slowest convergence occurring for $|\zeta|$ close to $1$. Fig.~\ref{fig:zeta} shows a contour plot of $|\zeta|$ where
$h=g^{-1}$ which shows that the speed of convergence will be slower when $g$ is closer
to the real axis between $\left(-\infty, -1/4 \right)$ (corresponding to $h\in\left(-4, 0\right)$).

\begin{figure}[!tb]
  \centering
      \includegraphics[trim=0mm 10mm 0mm 10mm, clip, scale=0.3]{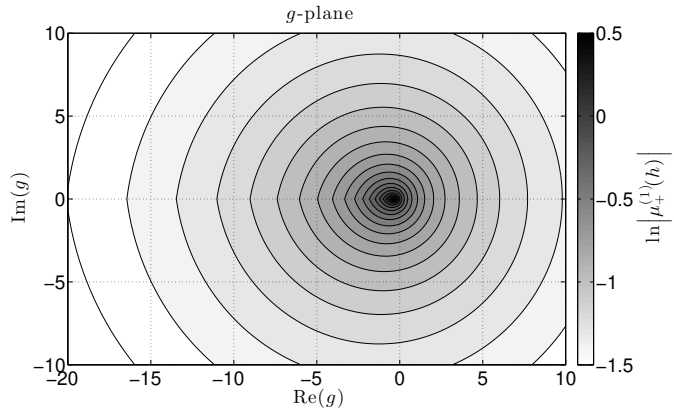}
    \caption{Contour plot of $|\mu_+^{(1)}(h)|$ where $h=g^{-1}$.}
    \label{fig:mu+}
\end{figure}

A contour map of the magnitude of $\mu_+^{(1)}$ in the $g$-plane is shown in Fig.~\ref{fig:mu+}.
The outermost boundary represents $\ln|\mu_+^{(1)}|=-1.5$ and the spacing of the contours is 0.1. As stated in
the remark of Theorem~\ref{thm:absmu}, $|\mu_+^{(1)}|$ takes its maximum value $2~(\approx \ln(0.693))$
when $g=-1/4$. The figure shows that the asymptotic value of $F_N^{(1)}(h(j\omega))$ as $N \rightarrow \infty$ is
directly related to the proximity of $h(j\omega)^{-1}$ to the point $-1/4$.

\begin{figure}[!tb]
  \centering
  \includegraphics[trim=0mm 32mm 0mm 30mm, clip, scale=0.3]{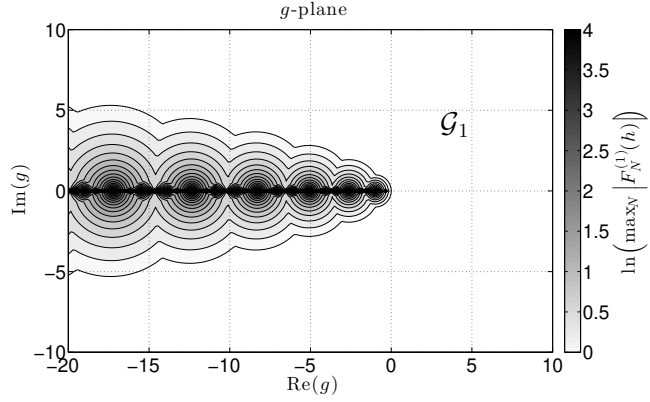}
  \caption{Contour plot of $\max_N|F_N^{(1)}(h)| = \gamma$ for $\ln(\gamma)=0, 0.2, 0.4, \dots$ where $h=g^{-1}$.}
   \label{fig:iterativemap}
\end{figure}

We now illustrate graphically the boundedness result of Theorem~\ref{thm:scalability}.
Fig.~\ref{fig:iterativemap} shows the region of the complex values of $g$~$(=h^{-1})$ for which
$\max_N|F_N^{(1)}(h)| \leq \gamma$ with $1\leq N \leq 200$ for a positive constant $\gamma$. The spacing of
the contours is 0.2 where $\ln(\gamma)$ takes the value $0, 0.2, 0.4, \dots$. The outermost boundary
represents $\gamma = 1$ and $\mathcal{G}_1$ denotes the set $\{g\in \mathbb{C} : \max_N|F_N^{(1)}(g^{-1})| \leq 1\}$.
This means that $\max_N \|{F_N^{(1)}}(h(s))\|_\infty \leq 1$ if and only if $g(s) \in \mathcal{G}_1 \mbox{ for } s
\in \mathbb{C}_+$. Note that from Fig.~\ref{fig:zeta} the sequence $\{F_N^{(1)}\}$ converges to the fixed point
$\mu_+^{(1)}$ quickly when $g \in \mathcal{G}_1$ so the choice of $N=200$ is large enough to accurately
determine the shape of the boundary in the figure.

\begin{figure}[!tb]
  \centering
  \includegraphics[trim=10mm 45mm 0mm 40mm, clip, scale=0.3]{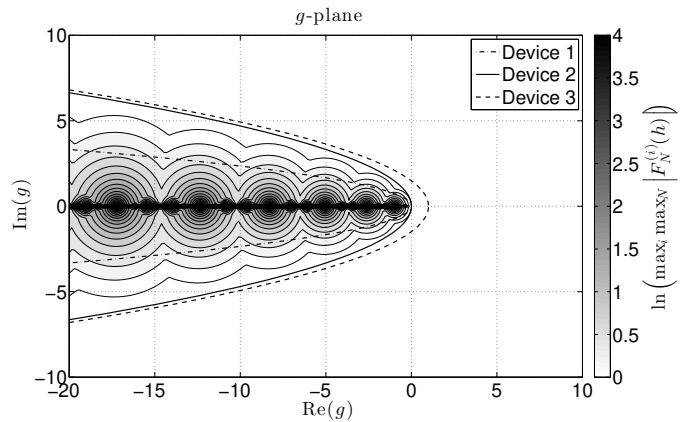}
  \caption{Nyquist diagrams of $g(s)=Y(s)/(sm)$ for the vibration control devices
                  in Table \ref{tab:exparameter} and contour plot of $\max_i\max_N|F_N^{(i)}(h)|=\gamma$
                for $\ln(\gamma)=0, 0.2, 0.4, \dots$ where $h=g^{-1}$.}
   \label{fig:pmapall}
\end{figure}

\begin{figure}[!tb]
 \centering
 \includegraphics[trim=15mm 0mm 15mm 5mm, clip, scale=0.3]{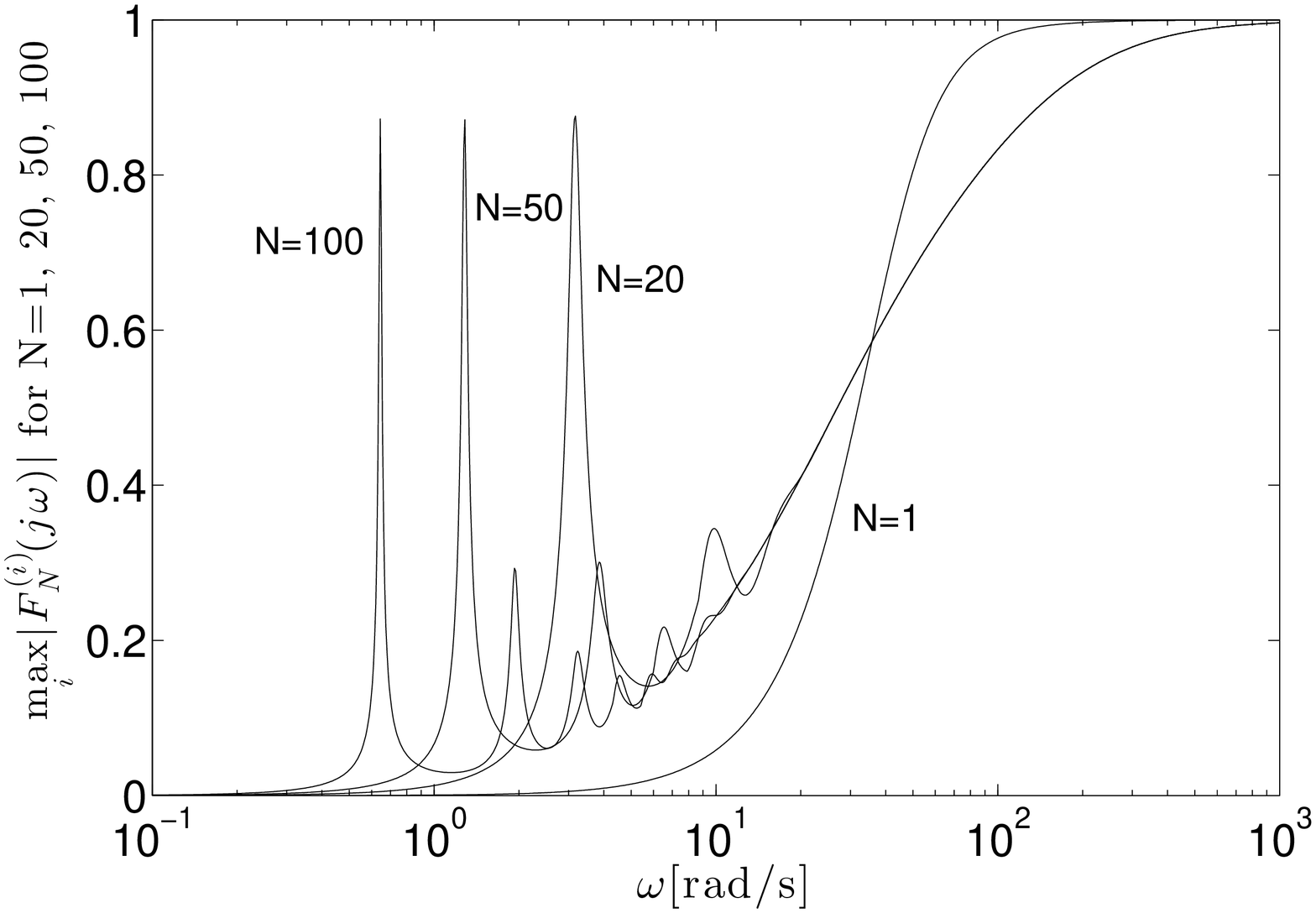}
  \caption{$\max_i|F_N^{(i)}(h(j\omega))|$ using Device 2 for $N=1, 20, 50, 100$.}
\label{fig:freqrespD2}
\end{figure}

\begin{figure}[!tb]
 \centering
 \includegraphics[trim=15mm 0mm 15mm 5mm, clip, scale=0.3]{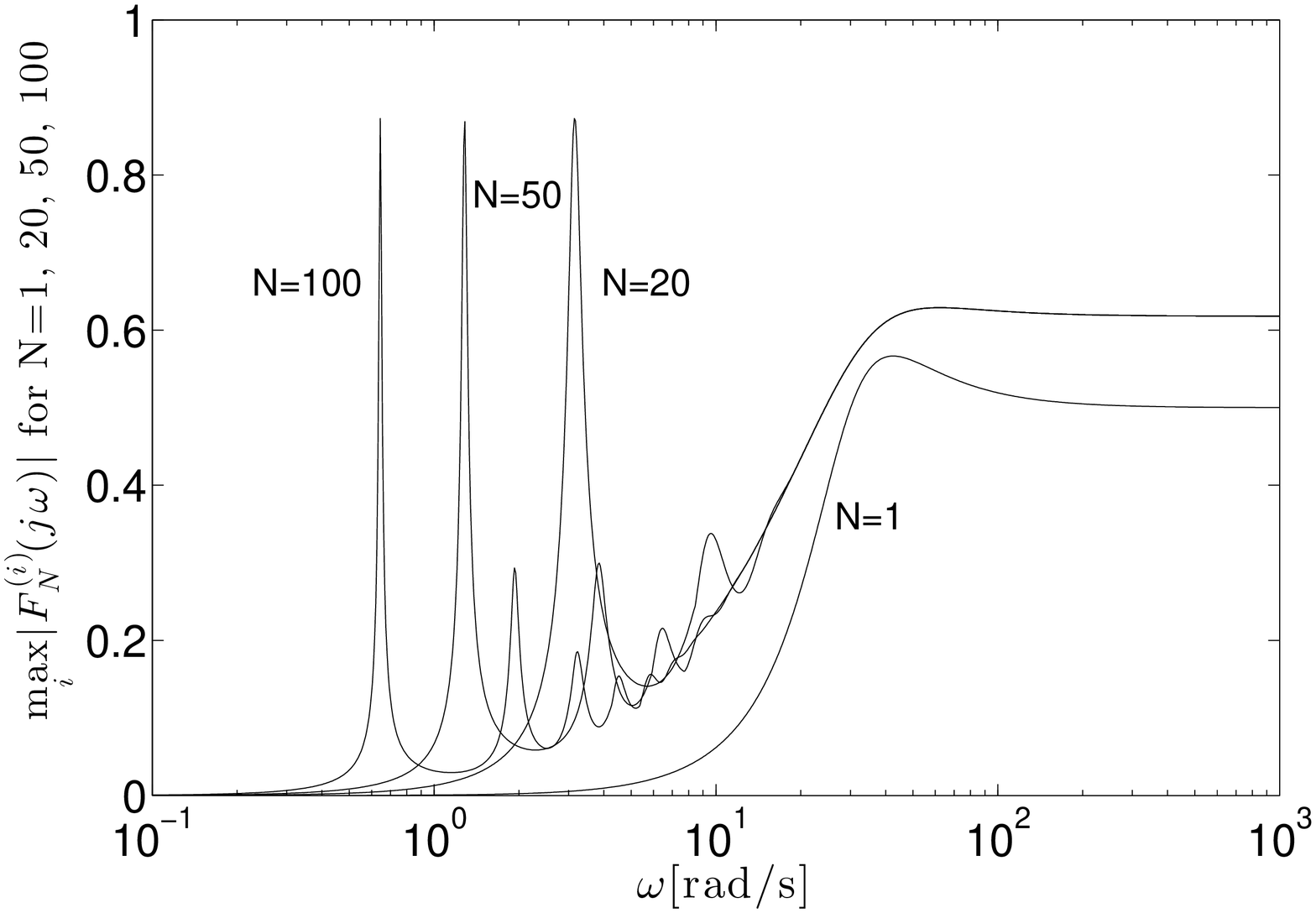}
  \caption{$\max_i|F_N^{(i)}(h(j\omega))|$ using Device 3 for $N=1, 20, 50, 100$.}
\label{fig:freqrespD3}
\end{figure}

\begin{figure}[!tb]
  \centering
  \includegraphics[trim=0mm 20mm 0mm 20mm, clip, scale=0.3]{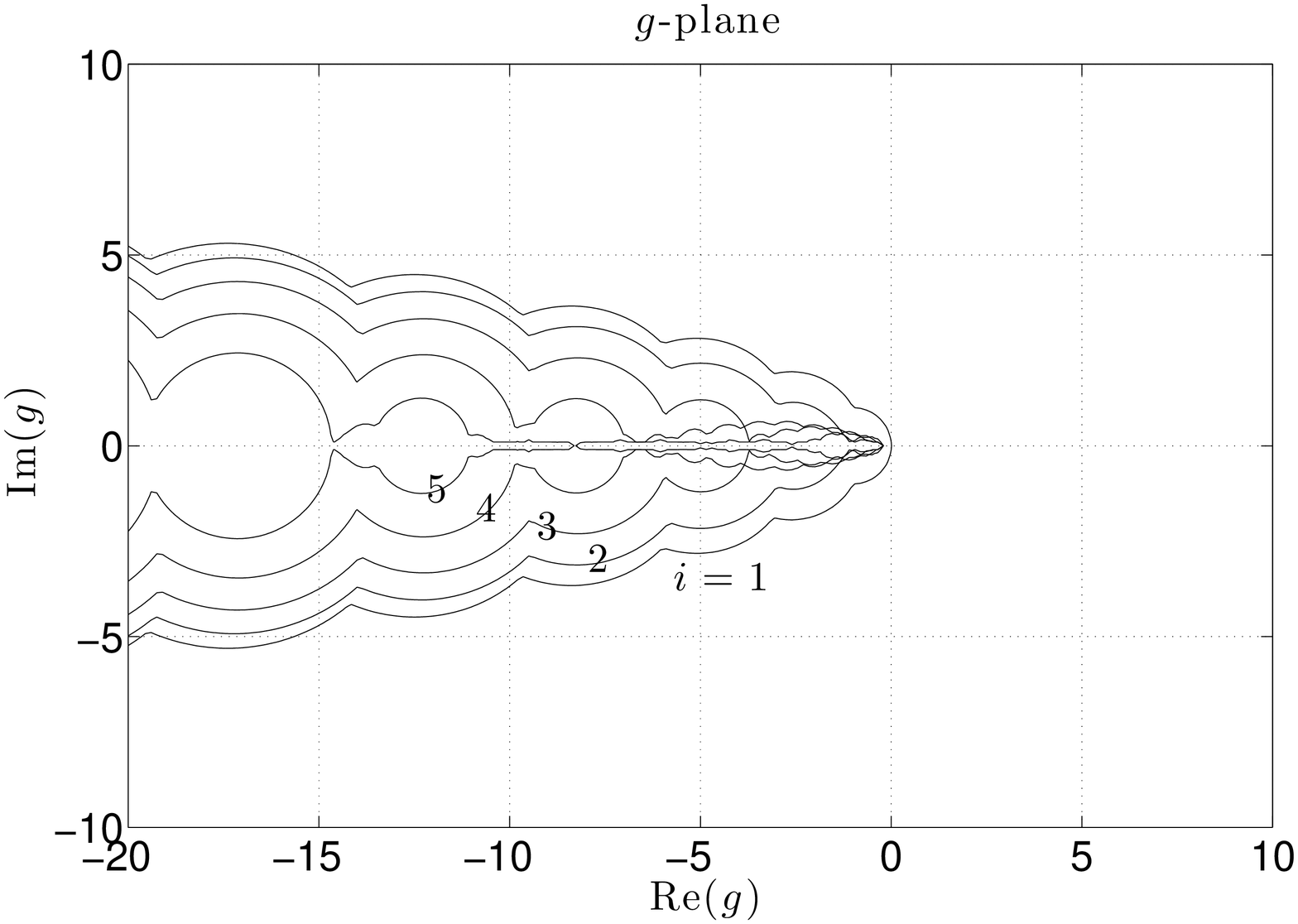}
  \caption{Curves representing $\max_N| F_N^{(i)}(h)|=1$ for $i=1, 2,\dots, 5$ where $h=g^{-1}$.}
   \label{fig:norm1}
\end{figure}

Fig.~\ref{fig:pmapall} is a similar figure to Fig.~\ref{fig:iterativemap} but shows a contour map of
$\max_i\max_N|F_N^{(i)}(h)| = \gamma  \in \mathbb{R}_+$ for $i=1, 2, \dots, N$, $i \leq N \leq 200$
with the Nyquist diagrams of $g(s)$ of three passive vibration control devices.
The layouts of these devices are shown in Table~\ref{tab:layouts} and their structural parameters are given in
Table~\ref{tab:exparameter}. We fix the parameters of the building model as $m=1.0\times10^5\mbox{ kg},
k=1.7\times10^5\mbox{ kN/m}$ (based on values given in \cite{Leg92}).
The outermost boundary of the contours again represents $\gamma = 1$ so
$\max_i \max_N \|{F_N^{(i)}}(h(s))\|_\infty \leq 1$ if the Nyquist diagram $g(j\omega)$ lies outside this boundary.
We see that devices 2 and 3 achieve this. It is also observed that the use of the
inerters improves the high frequency performance (corresponding to the origin in the $g$-plane).
The frequency domain plots of $\max_i|F_N^{(i)}(h(j\omega))|$ (Figs.~\ref{fig:freqrespD2} and \ref{fig:freqrespD3})
confirm these observations. Fig.~\ref{fig:norm1} shows the curves which represent $\max_N |F_N^{(i)}|=1$
where $i=1, 2,\dots, 5$ with $1\leq N \leq 200$. We observe that the set
$\{g\in \mathbb{C} : \max_N |F^{(1)}_N(g^{-1})| \leq 1\}$ contains the sets
$\{g\in \mathbb{C} : \max_N |F_N^{(i)}(g^{-1})| \leq 1\}$, $i=2, \dots, 5$.
\begin{table}[!tb]
 \centering
 \caption{Vibration control device layouts.}\vspace{-0.5cm}
 \label{tab:layouts}
 \begin{tabular}{c c c}\\
  \hline
  \vspace{-.1cm}&\\
  L1    &    L2 \vspace{.1cm}  \\ \hline
  \vspace{-.1cm}&\\
  \vspace{.3cm}$Y(s)= \displaystyle c+\frac{k}{s}$ & $Y(s)= \displaystyle bs+c+\frac{k}{s}$\\
  \includegraphics[trim=10mm 0mm 20mm 10mm, clip, scale=0.18]{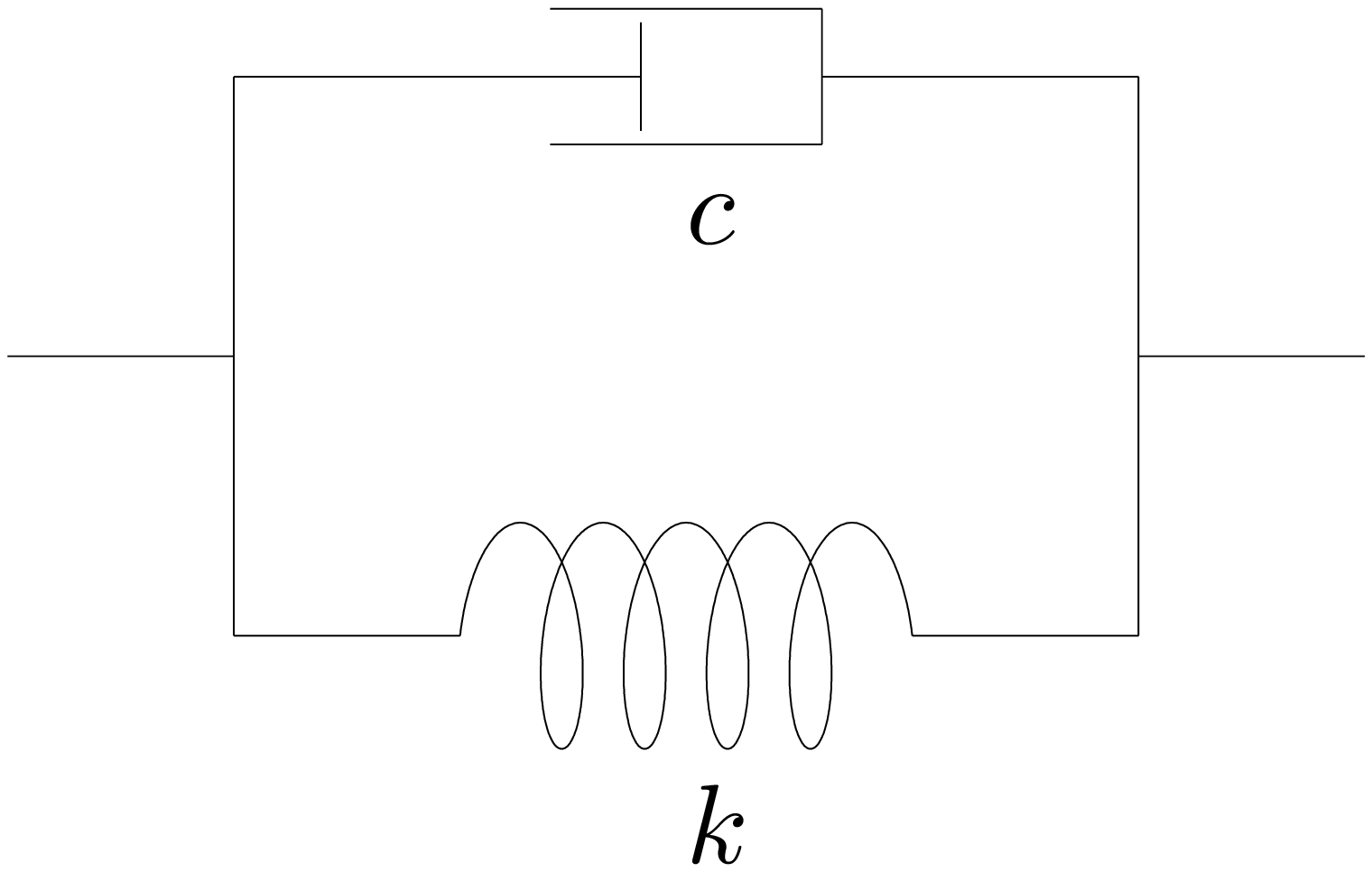}
  & \includegraphics[trim=10mm 0mm 20mm 10mm, clip, scale=0.18]{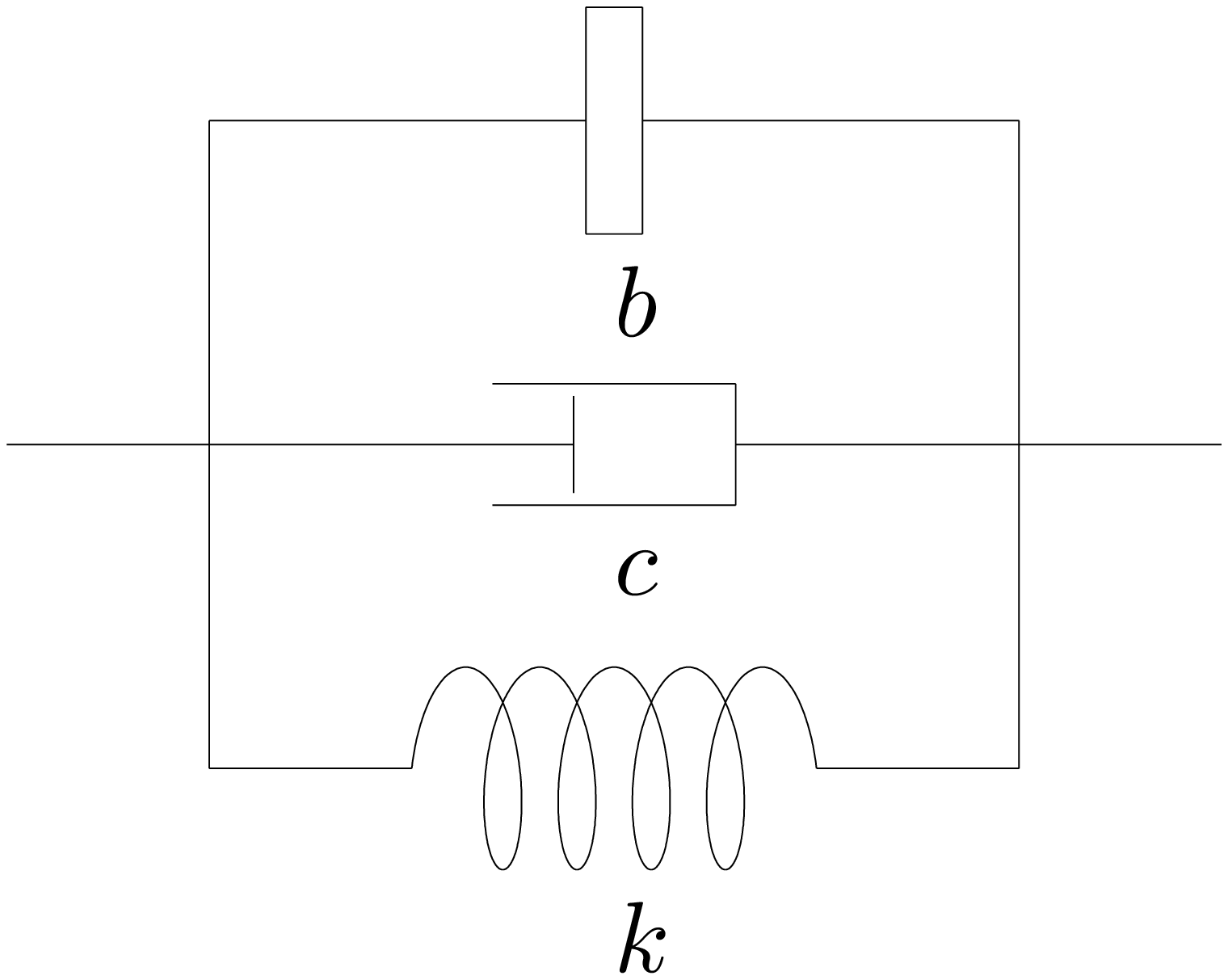} \\ \hline
 \end{tabular}
\end{table}
\begin{table}[!tb]
 \centering
 \caption{Parameters of vibration control devices.}\vspace{-0.5cm}
 \label{tab:exparameter}
 \begin{tabular}{@{\hspace{0.5cm}}l@{\hspace{.5cm}}c@{\hspace{.5cm}}
                                                    c@{\hspace{.5cm}}c@{\hspace{0.5cm}}}\\[0.2ex]
  \hline
  \vspace{-.3cm}&&&\\
        &  Layout & $c$ [kNs/m] & $b$ [kg] \\
  \vspace{-.3cm}&&&\\ \hline
  \vspace{-.2cm}&&&\\  \vspace{0.2cm}
    Device 1    &  L1 &  {$4.0\times10^3$} & --    \\  \vspace{0.2cm}
    Device 2    &  L1 &  {$6.0\times10^3$} &  --  \\  \vspace{0.2cm}
    Device 3    &  L2 &  {$6.0\times10^3$} & $1.0\times10^5$ \\  \vspace{0.2cm}
    \vspace{-.5cm}&&&\\ \hline
 \end{tabular}
\end{table}
\section{Conclusions} \label{sec:concl}
The interconnection of a chain of $N$ identical masses has been studied in which neighbouring masses are
connected by identical two-terminal passive mechanical impedances, and where the first mass is also connected
by the same impedance to a movable point. The problem is similar to that of symmetric bidirectional control of
a vehicle string, albeit with a passivity constraint. Formulae for the transfer functions from the movable point
displacement to a given intermass displacement have been derived in the form of complex iterative maps as a
function of a dimensionless parameter $h$ depending on the impedance and mass. The maps take the form of an
iterated M\"{o}bius transformation. It is shown that the fixed points of the mappings provide information on the
asymptotic behaviour of the disturbance transfer functions. Further, the use of a conjugacy transformation allows the
iterative map to be written in a convenient form to derive formal upper bounds on the infinity norm of the individual
transfer functions from the movable point displacement to a given intermass displacement. In particular they are
shown to be uniformly bounded with respect to $N$ for a suitable choice of $h$. This boundedness result was
illustrated graphically. The graph indicates the region of the complex plane for $h^{-1}$ to achieve a small infinity norm.
The method is presented in the context of the design of a multi-storey building. A comparison is made between a
standard spring-damper model for the lateral inter-storey suspension and the use of inerters.

%
\appendices

\section{Proof of Theorem~\ref{thm:ithrec}} \label{ap:thm:ithrec}
\begin{IEEEproof}
Define
\begin{align}
p(N,i)=&\overbrace{(d_{N-i+1}-d_{N-i})(d_{N-i}-d_{N-i-1}}+d_{N-1}d_i) \nonumber\\
        & - d_Nd_{i-2}(d_{N-i}-d_{N-i-1}) - hd_{N-1}d_N.  \label{eq:deff}
\end{align}
From \eqref{eq:TF} and the recursion of $F_N^{(i)}$ in \eqref{eq:ithrec},
we see that the theorem is equivalent to $p(N,i)=0$ for all $i \in \mathbb{N}$ and $i \leq N \in \mathbb{N}$.
The proof will follow by induction after establishing the following facts:
\begin{enumerate}
  \item $p(N,1)=0$ for all $N \geq 1$.
  \item $p(N,2)=0$ for all $N \geq 2$.
  \item $p(N, i) = p(N,i-1)+p(N-1,i-1)-p(N-1,i-2)$ for any $i \geq 3, N \geq i$.
\end{enumerate}
We now establish these facts in turn.
\begin{enumerate}
\item $\displaystyle
 \begin{aligned}[t]
p(N,1)={}&(d_{N}-d_{N-1} - d_{N}d_{-1})(d_{N-1}-d_{N-2})
              +d_{N-1}(d_1(d_{N}-d_{N-1}) - hd_N) \\
           ={}&d_{N-1}(d_N-(h+2)d_{N-1}+d_{N-2}) \\
           ={}&0,
 \end{aligned}$

 where the second step uses $d_{-1}=1$ and $d_1=h+1$, and the third step follows from \eqref{eq:dnrec}.
 \item $\displaystyle
 \begin{aligned}[t]
p(N,2)={}&(d_{N-1}-d_{N-2} - d_{N}d_0)(d_{N-2}-d_{N-3})+d_{N-1}d_2(d_{N-1}-d_{N-2}) - hd_{N-1}d_N \\
           ={}&(- (h+1)d_{N-1})(d_{N-1}-(h+1)d_{N-2})+d_{N-1}(h^2+3h+1)(d_{N-1}-d_{N-2})-hd_{N-1}d_N \\
           ={}&-hd_{N-1}(d_N-(h+2)d_{N-1}+d_{N-2})\\
            ={}&0,
 \end{aligned}$

 where the second step follows from $d_{N-2}-d_{N-3}=d_{N-2}-\left((h+2)d_{N-2}-d_{N-1}\right)$,
$d_{N-1}-d_{N-2} - d_{N}d_0=d_{N-1}-\left((h+2)d_{N-1}-d_{N}\right)-d_N$ and $d_2=h^2+3h+1$ using \eqref{eq:dnrec}.
\item
Consider the expression
\begin{equation}
X(N,i) = p(N,i) - p(N,i-1) -p(N-1,i-1)+ p(N-1,i-2). \label{eq:ind}
\end{equation}
It may be observed that four terms in \eqref{eq:ind} corresponding to the overbrace in \eqref{eq:deff}
cancel pairwise. Also the four terms in \eqref{eq:ind} of the form $-hd_{N-1}d_N$ cancel pairwise.
Thus
\begin{align}
X(N,i)={}&(d_{N-i}-d_{N-i-1})(d_{N-1}d_{i-3}-d_{N}d_{i-2}) \nonumber\\
            &+(d_{N-i+1}-d_{N-i})(d_{N-1}d_{i}+d_{N}d_{i-3} -d_{N-2}d_{i-1}-d_{N-1}d_{i-4}) \nonumber\\
           & +(d_{N-i+2}-d_{N-i+1})(d_{N-2}d_{i-2}-d_{N-1}d_{i-1}). \label{eq:ind2}
 \end{align}
Using the following substitution
\begin{align*}
&d_i=(h+2)d_{i-1}-d_{i-2} \\
&d_N=(h+2)d_{N-1}-d_{N-2} \\
&d_{i-1}=(h+2)d_{i-2}-d_{i-3} \\
&d_{i-4}=(h+2)d_{i-3}-d_{i-2}
\end{align*}
in the second term of \eqref{eq:ind2} and rearranging gives
\begin{align*}
X(N,i)={}&(d_{N-i}-d_{N-i-1})(d_{N-1}d_{i-3}-d_{N}d_{i-2})\\
            &+(d_{N-2}d_{i-2}-d_{N-1}d_{i-1})(d_{N-i+2}-d_{N-i+1}-(h+2)(d_{N-i+1}-d_{N-i})).
\end{align*}
Now note that
\begin{align*}
     d_{N-i+2}-d_{N-i+1}-(h+2)(d_{N-i+1}-d_{N-i})&=(h+1)d_{N-i}-d_{N-i+1} \\
     &=-(d_{N-i}-d_{N-i-1}).
\end{align*}
Hence
\begin{align*}
X(N,i)={}&(d_{N-i}-d_{N-i-1})(d_{N-1}d_{i-3}-d_{N}d_{i-2}-d_{N-2}d_{i-2}+d_{N-1}d_{i-1})\\
          ={}&(d_{N-i}-d_{N-i-1})(d_{N-1}d_{i-3}-(h+2)d_{N-1}d_{i-2}\\
          &+d_{N-2}d_{i-2}-d_{N-2}d_{i-2}+(h+2)d_{N-1}d_{i-2}-d_{N-1}d_{i-3})\\
          ={}&0.
 \end{align*}
\end{enumerate}
\end{IEEEproof}

\section{Proof of \eqref{eq:multiplier} and \eqref{eq:zetaquad}}  \label{ap:zetaellipse}
\begin{IEEEproof}
First we show \eqref{eq:multiplier}.
\begin{align*}
\frac{d_{i-2}-\mu_+^{(i)}}{d_{i-2}-\mu_-^{(i)}} &= \frac{d_{i-2}-\mu_+^{(i)}}{d_{i-2}-\mu_-^{(i)}}\times \frac{d_{i-2}-\mu_+^{(i)}}{d_{i-2}-\mu_+^{(i)}} \\
  &= \frac{\left(d_{i-2}-\mu_+^{(i)}\right)^2}{d_{i-2}^2 - d_{i-2}\left(\mu_+^{(i)}+\mu_-^{(i)}\right)+\mu_+^{(i)}\mu_-^{(i)}} \\
  &= \frac{\left(d_{i-2}-\mu_+^{(i)}\right)^2}{d_{i-2}d_i-h} \\
  &=\left( \frac{d_{i-2}-\mu_+^{(i)}}{d_i}\right)^2 \\
  &=\zeta^2
\end{align*}
where the third step follows from $\mu_+^{(i)}+\mu_-^{(i)} = -(d_{i} - d_{i-2})$ and $\mu_+^{(i)}\mu_-^{(i)} = -h$ \big(see \eqref{eq:quadithmu}\big)
and the fourth step follows from \eqref{eq:disquared}.

To show \eqref{eq:zetaquad}, note that
\begin{align*}
\zeta+\frac{1}{\zeta} &= \frac{d_{i-2}-\mu_+^{(i)}}{d_{i-1}}+ \frac{d_{i-1}}{d_{i-2}-\mu_+^{(i)}} \\
 &=\frac{d_{i-2}^2-2d_{i-2}\mu_+^{(i)} + {\mu_+^{(i)}}^2 + d_{i-1}^2}{d_{i-1}(d_{i-2}-\mu_+^{(i)})} \\
 &=\frac{(d_{i-2}+d_i)(d_{i-2}-\mu_+^{(i)})}{d_{i-1}(d_{i-2}-\mu_+^{(i)})} \\
 &=h+2.
\end{align*}
where the third step substitutes for ${\mu_+^{(i)}}^2$ and $d_{i-1}^2$ from \eqref{eq:quadithmu} and \eqref{eq:disquared} and
the fourth step uses \eqref{eq:dnrec}.
\end{IEEEproof}

\section{Proof of $\mu_+^{(i)} / \mu_-^{(i)} = -\zeta^{2i-1}$} \label{ap:mupovermun}
\begin{IEEEproof}
Since $\mu_\pm^{(i)}$ are the solutions of \eqref{eq:quadithmu}, $\mu_+^{(i)}\mu_-^{(i)} = -h$.
Therefore,
\begin{align*}
\frac{\mu_+^{(i)}}{\mu_-^{(i)}} &=-\frac{{\mu_+^{(i)}}^2}{h} \\
&=-\frac{\left(\zeta^{i-1}\mu_+^{(1)}\right)^2}{h}\\
&=-\zeta^{2i-2}\left(1-\mu_+^{(1)}\right) \\
&=-\zeta^{2i-1},
\end{align*}
where we use \eqref{eq:murec}, \eqref{eq:quadithmu} and \eqref{eq:zeta} in the second, third and fourth steps
above.
\end{IEEEproof}


\section{Proof of Theorem~\ref{thm:scalability}} \label{ap:scalability}
\begin{figure}[h]
\centering
   \begin{overpic}[trim=0mm 20mm 0mm 0mm, clip,scale=0.4]{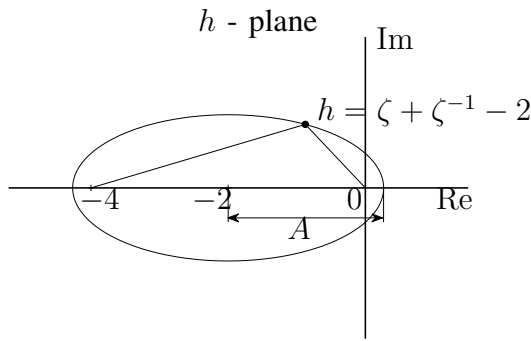}
   \put(60,20){$A$}
   \put(25,25){$-4$}
    \put(70,25){$0$}
    \put(44,25){$-2$}
   \put(65,40){$h=\zeta+\zeta^{-1}-2$}
   \put(75,52){$\operatorname{Im}$}
   \put(85,25){$\operatorname{Re}$}
   \put(45,55){$h$ - plane}
  \end{overpic}
   \caption{The ellipse defined by $h=\zeta+\zeta^{-1}-2$ for $|\zeta|$ fixed and
   $\mathrm{arg}(\zeta)$ varying.}\label{fig:circle2ellipse}
 \end{figure}

We first establish two lemmas which we need in the proof of this theorem.
The first lemma gives a relatively straightforward bound on $\sup_{N\geq i}|F_N^{(i)}(h)|$ for $h$
bounded away from the interval $[-4, 0]$. The second lemma is significantly more delicate and
deals with the fact that $h(j\omega) \rightarrow 0$ as $\omega \rightarrow 0$. The manner in which
this convergence occurs is critical to establish an upper bound.
\begin{Lemma} \label{lem:ellipsebound}
Let $h\in \mathbb{C}$ lie on an ellipse with foci $(-4, 0)$ and $(0, 0)$, and semi-major axis $A$.
Then
\begin{equation*}
\sup_{N\geq i}|F_N^{(i)}|  \leq \frac{(1+|\zeta|)(1+|\zeta|^2)}{1-|\zeta|^3}
\end{equation*}
where $\zeta$ is defined by \eqref{eq:zeta}. Moreover
\begin{equation*}
|\zeta|=\displaystyle\frac{A-\sqrt{A^2-4}}{2}, \quad A= \displaystyle\frac{|h|+|h+4|}{2}.
\end{equation*}
\end{Lemma}
\begin{IEEEproof}
From \eqref{eq:zetaquad}, $\zeta+\zeta^{-1}=h+2$. Letting ${\zeta = |\zeta| \mathrm{e}^{j\theta}},~ {-\pi} \leq \theta \leq \pi$, we have
\begin{equation} \label{eq:ellipsetransRandI}
\left(|\zeta|+\frac{1}{|\zeta|}\right) \cos\theta + j\left(|\zeta|-\frac{1}{|\zeta|}\right) \sin\theta = h+2.
\end{equation}
Keeping $|\zeta|$ fixed and solving \eqref{eq:ellipsetransRandI} for $h$ as a function of $\theta$ gives an ellipse
in the $h$-plane with centre $(-2, 0)$, foci $(-4, 0)$, $(0, 0)$ and semi-major axis $A=|\zeta| + 1/ |\zeta|$ (Fig.~\ref{fig:circle2ellipse}).
Therefore $|\zeta| = (A-\sqrt{A^2-4})/2$ where $A > 2$ and $|\zeta| < 1$.
Since the sum of the distances from the two foci and to a point on the ellipse is constant and equal to the major axis,
\begin{equation}\label{eq:semimajoraxis}
A= \frac{|h|+|h+4|}{2}.
\end{equation}
Then from \eqref{eq:mobF}, for all $N \geq i \geq 1$,
\begin{align*}
\sup_{N\geq i}|F_N^{(i)}| &\leq |\mu_+^{(1)}|\displaystyle\frac{1+|\zeta|^2}{1-|\zeta|^3} \\
                                        &\leq \frac{(1+|\zeta|)(1+|\zeta|^2)}{1-|\zeta|^3}
\end{align*}
since $|\mu_+^{(i)}| <|\mu_+^{(1)}|$ as shown in Theorem~\ref{thm:absmu} and $\mu_+^{(1)} = 1 - \zeta$.
\end{IEEEproof}
\begin{Lemma} \label{lem:limUB}
Let $h(s)$ be defined as in Theorem~\ref{thm:scalability}. Then $h(j\omega) = - c_1\omega^2 + jc_2\omega^3+\omega^4h_1(j\omega)$
for $\omega \geq 0 $ where $c_1$ and $c_2$ are positive constants and $|h_1(j\omega)| \leq c_3 \in \mathbb{R}_+$ for $0\leq \omega \leq \omega_1$.
Furthermore there exists $\omega_0$ with $0 < \omega_0 \leq \omega_1$ such that
\begin{equation*}
\sup_{N\geq i}|F_N^{(i)}(h(j\omega))|  < \dfrac{2\sqrt{c_1+c_2+c_3}\omega_0}
            {1- \exp{\left(-\displaystyle\frac{(c_2-\omega_0 c_3)\pi\omega_0}{8c_4^2}\right)}}
\end{equation*}
for $0 \leq \omega <  \omega_0$ where $c_4= \sqrt{2(2c_1+c_2+2c_3)}$.
\end{Lemma}
\begin{IEEEproof}
First note that when $\omega=0$, $h(j\omega)=0$ and $|F_N^{(i)}| = 0$ for any $N\geq i$ from \eqref{eq:ithrec}.

When $\omega \neq 0$, we see that $h\not\in [-4, 0]$ and therefore $\zeta$ as defined in \eqref{eq:zeta} is nonreal and $|\zeta| < 1$.
Hence $|1-\zeta^{2(N-i+1)}| < 1+|\zeta|^2$. The magnitude of the denominator in \eqref{eq:mobF} takes its smallest value when
${\zeta^{2N+1}}$ is at the closest point to $-1$. Let $p$ be the positive real number such that ${\zeta^p}$ has the minimum real part.
With $p$ defined in this way, the smallest value is always larger than ${1-|\zeta|^p}$. Therefore,
\begin{equation}
 \sup_{N\geq i}|F_N^{(i)}| \leq |\mu_+^{(1)}|\displaystyle\frac{1+|\zeta|^2}{1-|\zeta|^p}. \label{eq:UB}
\end{equation}
 Since ${\mu_+^{(1)}}^2 = h(1-\mu_+^{(1)}) = h\zeta $ from \eqref{eq:quadithmu} and \eqref{eq:zeta},
for $0 < \omega \leq \min\{1, \omega_1\}$,
\begin{align}
\left|\mu_+^{(1)}\right|^2    & < |h| \nonumber\\
					& \leq |-c_1\omega^2+jc_2\omega^3| + |\omega^4h_1(j\omega)| \nonumber\\
					& \leq \sqrt{(c_1^2+c_2^2)\omega^4} + |h_1(j\omega)|\omega^2 \nonumber\\
					& \leq(c_1+c_2+c_3)\omega^2. \label{eq:modhub}
\end{align}
Therefore, the numerator in \eqref{eq:UB} is bounded above:
\begin{equation*}
|\mu_+^{(1)}|(1+|\zeta|^2) < 2\sqrt{c_1+c_2+c_3}\omega.
\end{equation*}

We next show a lower bound on the denominator in \eqref{eq:UB}.
First note the general inequality
\begin{equation} \label{eq:semimajor}
|\zeta| +\frac{1}{|\zeta|} > 2.
\end{equation}
Defining $A$ as in \eqref{eq:semimajoraxis}, we obtain $A\leq 2+|h|$.
Writing ${\zeta = |\zeta| \mathrm{e}^{j\theta}}$ we have, from \eqref{eq:zetaquad}, $A\cos \theta =\operatorname{Re}(h+2)
={2-(c_1-\omega^2\operatorname{Re}(h_1(j\omega)))\omega^2} \geq 2-(c_1+c_3)\omega^2$.
Hence,
\begin{align*}
\cos \theta &=\frac{\operatorname{Re}(h+2)}{A} \\
		&\geq \frac{2-(c_1+c_3)\omega^2}{2+|h|}= 1-\frac{(c_1+c_3)\omega^2+|h|}{2+|h|}
\end{align*}
and if $\omega \leq \sqrt{2/(c_1+c_3)}$, $-\pi/2 \leq \theta \leq \pi/2$.
On the other hand, $\cos \theta \leq 1- \theta^2/4$ when $-\pi/2 \leq \theta \leq \pi/2$.
Therefore, for $0 < \omega \leq \min \{1, \omega_1, \sqrt{2/(c_1+c_3)} \}$,
\begin{align*}
 \left|\theta\right| &\leq 2\sqrt{\displaystyle\frac{(c_1+c_3)\omega^2+|h|}{2+|h|}} \\
 			&\leq 2\sqrt{\displaystyle\frac{(c_1+c_3)\omega^2+(c_1+c_2+c_3)\omega^2}{2}} \\
			&= \sqrt{2(2c_1+c_2+2c_3)}\omega \\
			&=: c_4 \omega
\end{align*}
using \eqref{eq:modhub}.
Also, $\operatorname{Im}(h+2) = (c_2 + \omega \operatorname{Im}(h_1(j\omega)))\omega^3 \geq (c_2-\omega c_3)\omega^3 > 0$ when $\omega<c_2/c_3$. Hence
\begin{equation} \label{eq:semiminor}
\frac{1}{|\zeta|} - |\zeta| = \left|\frac{\operatorname{Im}(h+2)}{\sin\theta}\right| \geq \frac{(c_2-\omega c_3)\omega^2}{c_4}
\end{equation}
for $0 < \omega < \min \{1, \omega_1, \sqrt{2/(c_1+c_3)} , c_2/c_3\}$. We also note that $-\pi <\theta <0$ follows from \eqref{eq:ellipsetransRandI}
when $\operatorname{Im}(h+2)>0$.
Adding \eqref{eq:semiminor} to \eqref{eq:semimajor} gives
\begin{equation*}
\frac{2}{|\zeta|} >  2 + \frac{(c_2-\omega c_3)\omega^2}{c_4}
\end{equation*}
and therefore,
\begin{align} \label{eq:ineqzeta}
|\zeta|& < \frac{2c_4}{2c_4+(c_2-\omega c_3)\omega^2} \nonumber \\
          & \leq 1-\frac{(c_2-\omega c_3)\omega^2}{4c_4}
\end{align}
if $0 \leq \left(c_2-\omega c_3)\omega^2\right/2c_4 \leq 1$. This condition is satisfied if
$\omega \leq c_2/c_3$ and $\omega \leq \sqrt{2c_4/c_2}$.

Now let $\omega_0=\min\{1, \omega_1, \sqrt{2/(c_1+c_3)}, c_2/c_3, \sqrt{2c_4/c_2} \}$.
Since $p$ is the positive real number such that $\zeta^p$ has minimum real part, we see that $-\pi \leq p\theta <0$
on noting that $-\pi/2 \leq \theta <0$ for $\omega < \omega_0$.
We further observe that $-\pi \leq p\theta \leq -\pi/2$, which gives
\begin{equation}
p \geq \frac{\pi}{2|\theta|} > \frac{\pi}{2c_4\omega}. \label{eq:ineqp}
\end{equation}
Using \eqref{eq:ineqzeta} and \eqref{eq:ineqp}, since $\left(1-x/n \right)^n < e^{-x}$ for $x/n \leq 1$,
\begin{align*}
|\zeta|^p & <  \left(1-\frac{(c_2-\omega c_3)\omega^2}{4c_4} \right)^{\tfrac{\pi}{2c_4\omega}} \\
	&< \exp{\left(-\frac{(c_2-\omega c_3)\pi\omega}{8c_4^2}\right)} \\
	& < \exp{\left(-\frac{(c_2-\omega_0 c_3)\pi\omega}{8c_4^2}\right)}
\end{align*}
for $\omega < \omega_0$ which establishes the required lower bound on the denominator in \eqref{eq:UB}.

The proof is now complete since $ax/(1-e^{-bx})$ is a monotonically increasing function if $a$ and $b$ are positive.
\end{IEEEproof}

We will now prove Theorem~\ref{thm:scalability}.

\begin{IEEEproof}[Proof of Theorem~\ref{thm:scalability}]
For $h(j\omega)=mj\omega Z(j\omega)$, from Lemma~\ref{lem:limUB}, there exists $\omega_0 > 0$ such that
\begin{equation*}
\sup_{N\geq i}|F_N^{(i)}(h(j\omega))|  < \dfrac{2\sqrt{c_1+c_2+c_3}\omega_0}
            {1- \exp{\left(-\dfrac{(c_2-\omega_0 c_3)\pi\omega_0}{8c_4^2}\right)}}
\end{equation*}
for $0 \leq \omega <  \omega_0$ where $c_1, c_2, c_3$ and $c_4$ are positive constants defined in Lemma~\ref{lem:limUB}.
For $\omega_0 \leq \omega \leq \infty$, since $h(j\omega)$ is bounded away from $[-4, 0]$, from Lemma~\ref{lem:ellipsebound},
\begin{equation*}
\sup_{N\geq i}|F_N^{(i)}(h(j\omega))|  \leq \frac{(1+|\zeta_0|)(1+|\zeta_0|^2)}{1-|\zeta_0|^3}
\end{equation*}
where
\begin{equation*}
|\zeta_0|=\dfrac{A_0-\sqrt{A_0^2-4}}{2},  A_0=\min_{\omega \geq \omega_0} \left(\dfrac{|h(j\omega)|+|h(j\omega)+4|}{2}\right).
\end{equation*}
From Theorem~\ref{thm:stability}, $T_{\!{\hat{x}_0\rightarrow\hat{\delta}_i}}(=-F_N^{(i)})$ is a stable transfer function under the condition of Theorem~\ref{thm:scalability}.
Therefore the maximum modulus principle can be applied to complete the proof.
\end{IEEEproof}


\ifCLASSOPTIONcaptionsoff
  \newpage
\fi



\bibliographystyle{IEEEtran}
\bibliography{IEEEabrv,BibTAC14}

\begin{thebibliography}{10}
\providecommand{\url}[1]{#1}
\csname url@samestyle\endcsname
\providecommand{\newblock}{\relax}
\providecommand{\bibinfo}[2]{#2}
\providecommand{\BIBentrySTDinterwordspacing}{\spaceskip=0pt\relax}
\providecommand{\BIBentryALTinterwordstretchfactor}{4}
\providecommand{\BIBentryALTinterwordspacing}{\spaceskip=\fontdimen2\font plus
\BIBentryALTinterwordstretchfactor\fontdimen3\font minus
  \fontdimen4\font\relax}
\providecommand{\BIBforeignlanguage}[2]{{%
\expandafter\ifx\csname l@#1\endcsname\relax
\typeout{** WARNING: IEEEtran.bst: No hyphenation pattern has been}%
\typeout{** loaded for the language `#1'. Using the pattern for}%
\typeout{** the default language instead.}%
\else
\language=\csname l@#1\endcsname
\fi
#2}}
\providecommand{\BIBdecl}{\relax}
\BIBdecl

\bibitem{Soo97}
T.~T. Soong and G.~F. Dargush, \emph{Passive Energy Dissipation Systems in
  Structural Engineering}.\hskip 1em plus 0.5em minus 0.4em\relax Wiley New
  York, 1997.

\bibitem{Con98}
M.~C. Constantinou, T.~T. Soong, and G.~F. Dargush, \emph{Passive Energy
  Dissipation Systems for Structural Design and Retrofit}.\hskip 1em plus 0.5em
  minus 0.4em\relax Multidisciplinary Center for Earthquake Engineering
  Research Buffalo, New York, 1998.

\bibitem{Tak09}
I.~Takewaki, \emph{Building Control with Passive Dampers: Optimal
  Performance-based Design for Earthquakes}.\hskip 1em plus 0.5em minus
  0.4em\relax Wiley, 2009.

\bibitem{Smi02}
M.~C. Smith, ``Synthesis of mechanical networks: the inerter,'' \emph{{IEEE}
  Trans. Automat. Contr.}, vol.~47, no.~10, pp. 1648--1662, 2002.

\bibitem{Wan07}
F.~C. Wang, C.~W. Chen, M.~K. Liao, and M.~F. Hong, ``Performance analyses of
  building suspension control with inerters,'' in \emph{Proc. 46th {IEEE} Conf.
  Decision Control}, 2007, pp. 3786--3791.

\bibitem{Wan10}
F.~C. Wang, M.~F. Hong, and C.~W. Chen, ``Building suspensions with inerters,''
  \emph{Proc. Inst. Mech. Eng. C, J. Mech. Eng. Sci.}, vol. 224, no.~8, pp.
  1605--1616, 2010.

\bibitem{Laz14}
I.~Lazar, S.~Neild, and D.~Wagg, ``Using an inerter-based device for structural
  vibration suppression,'' \emph{Earthquake Engng. Struct. Dyn.}, vol.~43, pp.
  1129--1147, July 2014.

\bibitem{Swa96}
D.~Swaroop and J.~Hedrick, ``String stability of interconnected systems,''
  \emph{{IEEE} Trans. Automat. Contr.}, vol.~41, no.~3, pp. 349--357, Mar 1996.

\bibitem{Sei04}
P.~Seiler, A.~Pant, and J.~K. Hedrick, ``{Disturbance propagation in vehicle
  strings},'' \emph{{IEEE} Trans. Automat. Contr.}, vol.~49, no.~10, pp.
  1835--1841, Oct. 2004.

\bibitem{Bar05}
P.~Barooah and J.~P. Hespanha, ``{Error amplification and disturbance
  propagation in vehicle strings with decentralized linear control},'' in
  \emph{Proc. 44th {IEEE} Conf. Decision Control}, Seville, Spain, Dec 2005,
  pp. 4964--4969.

\bibitem{Les07}
I.~Lestas and G.~Vinnicombe, ``{Scalability in heterogeneous vehicle
  platoons},'' in \emph{Proc. Amer. Control Conf.}, New York, Jul. 2007, pp.
  4678--4683.

\bibitem{Mid10}
R.~H. Middleton and J.~H. Braslavsky, ``String instability in classes of linear
  time invariant formation control with limited communication range,''
  \emph{{IEEE} Trans. Automat. Contr.}, vol.~55, no.~7, pp. 1519--1530, 2010.

\bibitem{Jov05}
M.~Jovanovic and B.~Bamieh, ``On the ill-posedness of certain vehicular platoon
  control problems,'' \emph{{IEEE} Trans. Automat. Contr.}, vol.~50, no.~9, pp.
  1307--1321, Sept 2005.

\bibitem{Bam12}
B.~Bamieh, M.~Jovanovic, P.~Mitra, and S.~Patterson, ``Coherence in large-scale
  networks: dimension-dependent limitations of local feedback,'' \emph{{IEEE}
  Trans. Automat. Contr.}, vol.~57, no.~9, pp. 2235--2249, Sept 2012.

\bibitem{And06}
B.~D.~O. Anderson and S.~Vongpanitlerd, \emph{Network Analysis and Synthesis: A
  Modern Systems Theory Approach (Dover Books on Engineering)}.\hskip 1em plus
  0.5em minus 0.4em\relax Dover Publications, 2006.

\bibitem{Bru31}
O.~Brune, ``{Synthesis of a finite two-terminal network whose driving-point
  impedance is a prescribed function of frequency},'' \emph{J. Math. Phys.},
  vol.~10, pp. 191--236, 1931.

\bibitem{Val60}
M.~E.~V. Valkenburg, \emph{Introduction to Modern Network Synthesis}.\hskip 1em
  plus 0.5em minus 0.4em\relax Wiley, 1960.

\bibitem{Hor99}
R.~A. Horn and C.~R. Johnson, \emph{{Matrix Analysis}}.\hskip 1em plus 0.5em
  minus 0.4em\relax Cambridge University Press, 1999.

\bibitem{Har14}
S.~Hara, H.~Tanaka, and T.~Iwasaki, ``Stability analysis of systems with
  generalized frequency variables,'' \emph{{IEEE} Trans. Automat. Contr.},
  vol.~59, no.~2, pp. 313--326, 2014.

\bibitem{Dev89}
R.~L. Devaney, \emph{An Introduction to Chaotic Dynamical Systems},
  2nd~ed.\hskip 1em plus 0.5em minus 0.4em\relax Addison-Wesley, 1989.

\bibitem{Nee97}
T.~Needham, \emph{Visual Complex Analysis}.\hskip 1em plus 0.5em minus
  0.4em\relax Clarendon Press - Oxford University Press, 1997.

\bibitem{Bea01}
A.~F. Beardon, ``Continued fractions, discrete groups and complex dynamics,''
  \emph{Comput. Methods and Funct. Theory}, vol.~1, pp. 535--594, 2001.

\bibitem{Kno14}
S.~Knorn, A.~Donaire, J.~C. Ag{\"u}ero, and R.~H. Middleton, ``Passivity-based
  control for multi-vehicle systems subject to string constraints,''
  \emph{Automatica}, vol.~50, no.~12, pp. 3224 -- 3230, 2014.

\bibitem{Leg92}
P.~L{\'e}ger and S.~Dussault, ``Seismic-energy dissipation in {MDOF}
  structures,'' \emph{J. Struct. Eng.}, vol. 118, no.~5, pp. 1251--1269, 1992.

\end{thebibliography}
\end{document}